\documentclass[12pt]{article}
\usepackage{amsmath}
\usepackage{graphicx,psfrag,epsf}
\usepackage{enumerate}
\usepackage[left=1in,right=1in,top=1in,bottom=1in]{geometry}

\usepackage{setspace}
\usepackage{graphicx}
\usepackage{wrapfig}
\usepackage{epstopdf}
\usepackage{subcaption}
\usepackage{caption}
\usepackage{multirow}
\usepackage{nicefrac}
\usepackage{booktabs}
\usepackage{amsmath}
\usepackage{amsthm}
\usepackage[algo2e]{algorithm2e} 
\usepackage{verbatim} 
\usepackage{amssymb}
\usepackage{bbm}
\usepackage[english]{babel}
\usepackage{graphicx,epstopdf}
\RequirePackage[authoryear]{natbib}
\usepackage{hyperref}
\hypersetup{colorlinks=true,citecolor=blue}
\hypersetup{linkcolor=blue}
\usepackage{algorithm}
\usepackage[noend]{algpseudocode}
\usepackage{tikz}

\newcommand{\be}{\begin{eqnarray}}
\newcommand{\ee}{\end{eqnarray}}
\newcommand{\ba}{\begin{eqnarray*}}
\newcommand{\ea}{\end{eqnarray*}}

\usepackage{bm}

\usepackage{array}
\newcolumntype{L}[1]{>{\raggedright\let\newline\\\arraybackslash\hspace{0pt}}m{#1}}
\newcolumntype{C}[1]{>{\centering\let\newline\\\arraybackslash\hspace{0pt}}m{#1}}
\newcolumntype{R}[1]{>{\raggedleft\let\newline\\\arraybackslash\hspace{0pt}}m{#1}}

\usepackage{array}

\include{math-commands}

\newcommand{\footremember}[2]{%
    \footnote{#2}
    \newcounter{#1}
    \setcounter{#1}{\value{footnote}}%
}
\newcommand{\footrecall}[1]{%
    \footnotemark[\value{#1}]%
} 
\title{Link prediction for ex ante influence maximization on temporal networks}
\author{%
  Eric Yanchenko\footnote{Department of Statistics, North Carolina State University, Raleigh, North Carolina, ekyanche@ncsu.edu} \footremember{tokyo}{Department of Computer Science, Tokyo Institute of Technology, Ookayama, Tokyo, Japan}
  \and Tsuyoshi Murata\footrecall{tokyo}
  \and Petter Holme \footnote{Department of Computer Science, Aalto University, Espoo, Finland} \footnote{Center for Computational Social Science, Kobe University, Kobe, Japan}
  }

\begin{document}

\maketitle
\begin{abstract}
\noindent
Influence maximization (IM) is the task of finding the most important nodes in order to maximize the spread of influence or information on a network. This task is typically studied on static or temporal networks where the complete topology of the graph is known. In practice, however, the seed nodes must be selected before observing the future evolution of the network. In this work, we consider this realistic {\it ex ante} setting where $p$ time steps of the network have been observed before selecting the seed nodes. Then the influence is calculated after the network continues to evolve for a total of $T>p$ time steps. We address this problem by using statistical, non-negative matrix factorization and graph neural networks link prediction algorithms to predict the future evolution of the network and then apply existing influence maximization algorithms on the predicted networks. Additionally, the output of the link prediction methods can be used to construct novel IM algorithms. We apply the proposed methods to eight real-world and synthetic networks to compare their performance using the Susceptible-Infected (SI) diffusion model. We demonstrate that it is possible to construct quality seed sets in the {\it ex ante} setting as we achieve influence spread within 87\% of the optimal spread on seven of eight network. In many settings, choosing seed nodes based only historical edges provides results comparable to the results treating the future graph snapshots as known. The proposed heuristics based on the link prediction model are also some of the best-performing methods. These findings indicate that, for these eight networks under the SI model, the latent process which determines the most influential nodes may not have large temporal variation. Thus, knowing the future status of the network is not necessary to obtain good results for {\it ex ante} IM.
\end{abstract}


\clearpage

\section{Introduction}
Influence maximization (IM) is a canonical problem in the computational analysis of social networks where the goal is to find seed nodes that maximize the reach of information diffusion \citep{kempe2003maximizing, chen2009efficient, li2018influence}. Ever since Kempe {\it et al.}'s seminal paper \citep{kempe2003maximizing}, this topic has attracted great attention from researchers across various domains. The IM problem is typically studied under one of two settings. The first considers a static network where the information spreads through the network over time, but the topology of the network remains fixed. This assumption is violated in many real-world settings, so recently there has also been interest in developing methods where the network structure is also allowed to vary with time \citep{holme2012temporal}. But there are also unrealistic assumptions that typically accompany this case. For a given temporal network $\mathcal G = (G_1,\dots,G_T)$, it is typically assumed that the network topology is known for all $t\in\{1,\dots,T\}$, i.e., the {\it ex post} assumption. Then at time $t=1$, a researcher chooses a seed set to maximize the influence on the evolving network. In other words, the researcher can ``peer into the future'' to pick the most influential nodes at time $t=1$. Assuming {\it complete foresight} of the network evolution, however, is unrealistic. In practice, one must choose the seed nodes based on the past evolution of the network without knowing exactly what the network will look like in the future.  In the temporal network literature, this is known as the {\it ex ante} setting where the solution is based on forecasts and not actual results.

\begin{figure}
    \centering
    \includegraphics[width=\textwidth]{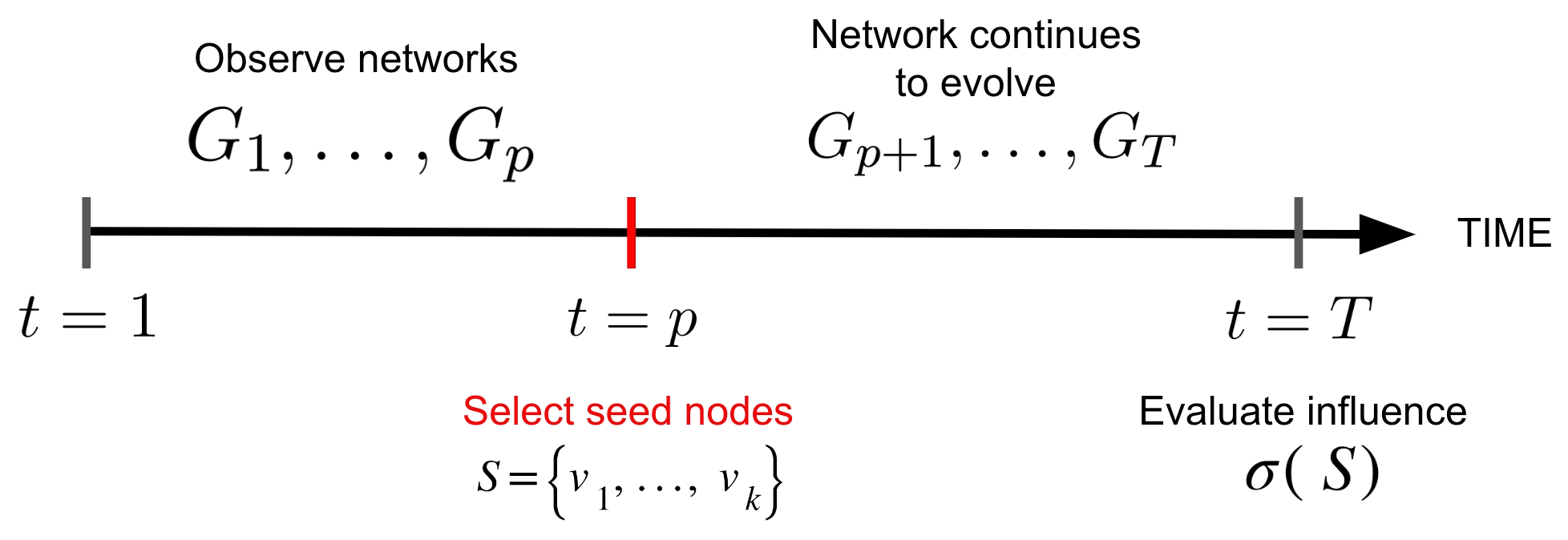}
    \caption{Schematic of proposed setting.}
    \label{fig:prob}
\end{figure}

In this paper, we consider the more realistic and difficult setting highlighted in Figure \ref{fig:prob}. Here, the researcher observes some network $G_1,\dots,G_p$ from $t=1,\dots,p$. Then after observing these first $p$ snapshots of the network, he/she chooses the seed nodes which will maximize the influence over the next $T-p$ snapshots, $G_{p+1},\dots, G_T$, {\it before} observing them. Then the number of influenced nodes after the spreading process has finished on the final snapshot is calculated. The formal problem statement is as follows: {\it given some network $\mathcal G=(G_1,\dots,G_p,\dots,G_T)$ evolving over time, what seed nodes should be chosen at time $t=p$ based only on $G_1,\dots,G_p$ in order to maximize the influence on the network at time $t=T+1$?}

\subsection{Influence maximization problem}
We begin by defining the influence maximization (IM) problem. Let $\mathcal G=(G_1,\dots,G_T)$ be a temporal network where the graph at time $t$, $G_t=(V,E_t)$ is a collection of nodes $V$ and edges $E_t$. Notice that the nodes are fixed but the edge set is allowed to vary over time. Each snapshot $G_t$ is defined by an $n\times n$ adjacency matrix $A^{(t)}$ such that $A_{ij}^{(t)}=1$ if nodes $i$ and $j$ have an edge at time $t$, and 0 otherwise. Given some process by which information is diffused on the network and integer $k<n$, the IM problem tries to find a set of nodes $S=\{v_1,\dots,v_k\}$ such that, if the nodes in $S$ are initially ``influenced'' or ``infected'' with the information, then the spread of influence on the network is maximized at time $t=T+1$.

Crucial to this problem is the information diffusion mechanism. There are many popular choices in the literature including the Independent Cascade (IC) model \citep{wang2012scalable, wen2017online}, Linear Threshold (LT) model \citep{chen2010scalable, goyal2011simpath}, triggering (TR) model \citep{kempe2003maximizing, tang2014influence}, and more. For this work, we adopt the susceptible-infected (SI) model often used in epidemiological studies \citep{allen1994some, murata2018extended}. At each step in the process\footnote{We assume that the graph evolution process and diffusion process run at the same time similar to \cite{gayraud2015diffusion}, i.e., the diffusion process occurs and then the graph evolves one time step before the diffusion process starts again.}, a node is either in a susceptible state (S)\footnote{Note that we use $S$ to represent the seed set and if a node is susceptible. From context, however, it should be clear which definition is used.} or infected state (I). If a node is in the S state, then the information has not yet reached it, and a node in the I state has already received the information. At the beginning of the process ($t=1$), nodes in the seed set are set to the I state and all other nodes are in the S state. Then at time $t$, if node $u$ is in I and node $v$ is in S and there is an edge between the two nodes, i.e., $A^{(t)}_{uv}=1$, then node $v$ will change to state I at time $t+1$ with probability $\lambda$. The susceptibility or infection parameter $\lambda$ controls the rate at which information propagates throughout the network. The diffusion process ends once $t=T+1$. Thus, if $\sigma(S)$ is the expected number of nodes infected at time $T+1$ when nodes in $S$ are initialized to I state under the SI model, the IM problem seeks the set of seed nodes of size $k$ that maximizes $\sigma(S)$ on $\mathcal G$, i.e.,
\begin{equation}
    S^*=\arg\max_{S\subseteq V, |S|=k}\sigma(S).
\end{equation}
The IM problem is NP-hard under the IC, LT, TR and SI models \citep[see, e.g.,][]{kempe2003maximizing, li2018influence}. Thus, the optimal solution is unfeasible in many cases so heuristics must be exploited to find a suitable seed set. Indeed, even evaluating $\sigma(S)$ is \#P-hard \citep[e.g.,][]{chen2010scalableaa, chen2010scalable}. In practice, we estimate $\sigma(S)$ via Monte Carlo (MC) simulations by simulating the spreading process a large number of times and taking the average number of nodes infected at time $T+1$.

Notice that in this definition, the future evolution of the network is taken as given. In other words, the seed nodes are selected at time $t=1$ under the assumption that the topology of the graph at time $t=2,\dots,T$ is known. In practice, however, this is an unreasonable assumption; a practitioner needs to select seed nodes {\it without} having access to the future dynamics of the network. While we are interested in this more realistic and difficult setting, we begin by highlighting some of the existing IM algorithms.

\subsection{Static IM}
We first highlight some of the existing approaches for IM in the static case, i.e., $T=1$. Please see \cite{li2018influence} for a comprehensive survey. \cite{kempe2003maximizing} was the first to postulate this as a combinatorial optimization problem and used a greedy algorithm to find the optimal seed set. At each step in the greedy algorithm, the node which maximizes the marginal gain in $\sigma(S)$ is added, and the process continues until $|S|=k$. Mathematically, node $v$ is added to $S$ where
\begin{equation}
    v=\arg\max_{u\in V\setminus S}\{\sigma(S\cup \{u\}) - \sigma(S)\}.
\end{equation}
The authors prove that the solution yielded by the greedy algorithm is within a factor of $(1-1/e)$ of the optimal solution. Thus, in studies it is often considered the ``gold-standard'' of IM. The approach, however, is extremely computationally expensive and, therefore, infeasible for large networks. Efforts to make this algorithm more efficient include estimating the upper bound on the marginal influence \citep{leskovec2007cost, goyal2011celf++} and simplifying the calculation of $\sigma(S)$ \citep{wang2010community}. Another class of IM algorithms ranks nodes according to some metric and then selects the $k$ nodes with largest metric value as the seed nodes. For example, in \cite{chen2009efficient}, nodes are ranked according to their degree and once a node $u$ is selected for the core, all nodes which share an edge with $u$ have their degree ``discounted'' by a specified factor to take into account the nodes' overlap in influence. \cite{liu2014influence} takes a similar approach for PageRank. These approaches avoid calculating $\sigma(S)$ which leads to computational speed-ups, but lack performance guarantees.

\subsection{Temporal IM}

Assuming that information diffuses on a static network is often times unreasonable, so recently there have been many works that study the IM problem for temporal or dynamic networks. First, the Greedy algorithm of Kempe can easily be extended to the temporal case and many extensions of this method have been studied \citep[e.g.,][]{liqing2019analysis, erkol2020influence}. \cite{michalski2014seed} show that temporal IM methods greatly outperform their static counterparts under the LT model. While this paper assumes that the future snapshots of the network are unknown, there is no attempt to forecast its evolution. \cite{osawa2015selecting} adopt the SI model and develop a heuristic method by approximating the probability that a node is infected in the next time step. This method performs similarly to a greedy algorithm but is significantly faster. Several static network IM heuristics are extended in \cite{murata2018extended} including a Dynamic Degree discount algorithm, based on dynamic degrees from \cite{yu2010finding}. This method is faster than greedy and \cite{osawa2015selecting} while yielding comparable influence spread. Lastly, \cite{erkol2020influence} leverage the SIR model and use an individual node mean field approximation to compute the expected influence with a greedy algorithm. One interesting finding from this work is that simply using the first temporal layer of the network to find the seed nodes can often times still yield good performance. This paper also briefly mentions the problem we are interested in, where the future evolution of the network is unknown, but do not thoroughly explore it.

\subsection{Main contributions}
In this paper, we study the temporal IM problem under the realistic {\it ex ante} setting where the future evolution of the network is unknown. Given the observed network snapshots, we first predict the future topology of the network using statistical, non-negative matrix factorization and graph neural network link prediction techniques. We also propose a novel heuristic for IM based on the link prediction model output. Then, we use greedy and dynamic degree IM algorithms to find the optimal seed nodes on the estimated future networks.  We conduct extensive experiments on synthetic and real-world networks and show that, in almost all cases, finding the optimal seed nodes on an aggregated graph using only the historical snapshots yields an influence spread within 80\% of the influence spread when the seed nodes are found using the actual future evolution. Additionally, the proposed IM heuristics yield influence spreads as good or better than actually predicting the future network evolution. These results together indicate the potential existence of an influential node latent process that does not vary temporally. More importantly, the IM problem can still be solved with good performance in this realistic and difficult setting. The remainder of this paper is structured as follows. In Section \ref{sec:methods}, we propose our methodology and discuss various link prediction and IM algorithms. We also propose a novel {\it ex ante} IM algorithm based on the link prediction models. We conduct experiments on seven real-world networks in Section \ref{sec:exp} and share concluding thoughts in Section \ref{sec:conc}.

Finally, the most similar works to ours consider a temporally evolving network and selecting a new seed set at each snapshot. For example, \cite{singh2021link} adopt the IC model for spreading dynamics and a conditionally temporal restricted Boltzmann machine (ctRBM) for link prediction \citep{li2014deep}. The authors propose to choose new seed nodes for each graph snapshot and update the set using an Interchange Heuristic \citep{nemhauser1978analysis}. Our paper differs from that of Singh's in several key areas including, but not limited to: only selecting a seed set once, predicting multiple time steps in the future, allowing edges to form and disappear, comparing several link prediction algorithms, and using the SI model. Our problem is also different than that of \cite{zhuang2013influence} and \cite{han2017influence}. These papers assume that the future networks are unobserved but that they can be partially known through probing different nodes. Conversely, our work assumes the future networks are completely unknown.

\section{Methodology}\label{sec:methods}
The goal of this paper is to develop a method for {\it ex ante} influence maximization where seed nodes are chosen before observing the future evolution of the network. We propose the following approach:
\begin{enumerate}
    \item Observe networks $G_1,\dots,G_p$.
    \item Predict the future evolution of the network $\hat G_{p+1},\dots,\hat G_T$ based on the observed networks.
    \item Apply an IM algorithm to the predicted networks $\hat G_{p+1},\dots,\hat G_T$ treating $t=p+1$ as the starting time to obtain optimal seed nodes $S=\{v_1,\dots,v_k\}$.
    \item Allow the network to continue to evolve as $G_{p+1},\dots,G_T$ and compute the influence of $S$ on the true networks.
\end{enumerate}
There are two key components in the proposed method: step (2) predicting the future networks and step (3) the IM algorithm.

\subsection{Link prediction}
The goal of link prediction methods is determining the most likely missing and/or future links in a network. There is a rich body of literature on this problem with numerous approaches such as statistical, non-negative matrix factorization (NMF) and graph neural networks (GNN). We refer the interested reader to the following surveys for a thorough review of link prediction methods \citep{lu2011link, kumar2020link, divakaran2020temporal, zhou2021progresses}. In general, static link prediction methods look for missing links in the network, thereby not allowing for the possibility of removing links. Temporal methods, on the other hand, must account for the possibility of new edges arising and existing edges disappearing. In our scenario, we are interested in the dynamic evolution of the network so we are interested in temporal link prediction methods. 

For this work, we consider one link prediction method from each of the following popular paradigms: statistical, NMF and GNN. 

\paragraph{Statistical:} \cite{zou2021temporal} use a linear regression model with LASSO penalty to predict future links in the network. We slightly modify their approach to perform logistic regression with LASSO. Let $x_i(t)=1$ if node pair $i$ had an edge at time $t$ and 0 otherwise, for $i=1,\dots,M$ where $M$ is the number of node pairs with at least one link in $G_1,\dots,G_p$. Then for $i=1,\dots,M$,
\begin{subequations}
\begin{equation}
    x_i(t+1)\sim\mathsf{Bernoulli}\left[\frac1{1+\exp\{-\nu_i(t)\}}\right]
\end{equation}
where
\begin{equation}
    \nu_i(t)=\beta_{i0}+\sum_{j=1}^M x_i(t)\beta_{ij}.
\end{equation}
\end{subequations}
Thus for each node pair $i$, we fit a logistic regression model to find the best fitting $\beta_i=(\beta_{i0},\beta_{i1},\dots,\beta_{iM})^T$. Since many edge pairs are likely uninformative in this model, the authors add a LASSO penalty ($L_1$ regularization) to shrink the absolute value of each $\beta_{ij}$ to 0, i.e., $\alpha\sum_{j=1}^M|\beta_{ij}|$ where $\alpha$ determines the strength of the penalty. A small value of $\alpha$ corresponds to little regularization and vice-versa for a large value. The optimal $\alpha$ is chosen from a grid of values to minimize the validations set area under the receiver operating curve (AUC). To predict the probability of an edge at several time steps in the future $\hat p_i(p+t)$, we sequentially use the fitted $\hat\beta_i$ and the estimated probabilities of an edge, i.e., 
\begin{subequations}
\begin{equation}
\hat p_i(p+t)=\text{expit}\{\hat\beta_{0i} + \sum_{j=1}^M \hat p_{i}(p+t-1)\hat\beta_{ij}\}
\end{equation}
where
\begin{equation}
\text{expit}(x)=\frac1{1+\exp(-x)}.
\end{equation}
\end{subequations}

The main advantage of this method is that it yields a valid probability of a link for each edge pair and not simply a similarity score like the following methods. We can also predict edges multiple time steps in the future without having to re-fit the model. A notable limitation of this method is that it can only predict links for edge pairs with at least one historical edge. Additionally, the simplicity of the linear model may not be able to capture the complex mechanism behind link formation and fitting separate models for every edge pair means that method may not scale well for networks with a large number of edges.

\paragraph{Non-negative matrix factorization:} For a given $n\times n$ matrix $A$, non-negative matrix factorization (NMF) seeks to find an $n\times q$ matrix $U$ and $q\times n$ matrix $V$ such that $A\approx UV$, $q<n$ and all entries of $U$ and $V$ are non-negative. In other words, $U$ and $V$ are low-dimensional representations of $A$. \cite{ahmed2018deepeye} applies this approach for link prediction in temporal networks. 
Let $A_1,\dots,A_p$ be the adjacency matrices corresponding to graphs $G_1,\dots,G_p$. Then the author seek to find a sequence of matrices $U_t,V_t$ such that $A_t\approx U_tV_t$ and all $U_t$ and $V_t$ are close to some consensus matrices $U_*$ and $V_*$, respectively. Mathematically, this means minimizing the following loss function:
\begin{subequations}
\begin{equation}\label{eq:nmf}
    L(U_t,V_t)
    =\sum_{t=1}^p \phi^{p-t}||A_t-U_tV_t||^2_F + \sum_{t=1}^p\phi^{p-t}||U_t-U_*||^2_F + \sum_{t=1}^p\phi^{p-t} ||V_t-V_*||^2_F
\end{equation}
where
\begin{equation}
    U_*
    =\frac{1}{M}\sum_{t=1}^t \phi^{p-t} U_t,\ \ \
    V_*
    =\frac{1}{M}\sum_{t=1}^t \phi^{p-t} V_t,\ \ \ 
    M
    =\sum_{t=1}^p \phi^{p-t},
\end{equation}
\end{subequations}
 subject to $U_t,V_t$ are non-negative. Here, $\phi$ is an attenuation coefficient gives a larger weight to more recent graphs and $||\cdot||_F$ is the Frobenius norm. The authors derive an iterative algorithm to minimize (\ref{eq:nmf}). Once the algorithm has converged, the the rows of $V_*$ represent a low-dimensional embedding for each node, $(V_*)_i$. Thus, $S_{ij}=\text{sim}\{(V_*)_i, (V_*)_j\}$ is a score for the likelihood of a link between nodes $i$ and $j$ at time $p+1$ where $(V_*)_k$ is the $k$th row of $V_*$ and $\text{sim}(\cdot,\cdot)$ is a measure of similarity. In this work, we use the cosine similarity. To extend this method to predict multiple time steps in the future, we first use $S_{ij}$ to predict $A_{t+1}$ using a threshold cutoff (see below for further discussion). Then the appropriate terms are added to the loss function to find the non-negative matrix factorization of this new adjacency matrix, i.e., 
\begin{equation}
    L(U_t,V_t)
    =\sum_{t=1}^{p+1} \phi^{p+1-t}||A_t-U_tV_t||^2_F + \sum_{t=1}^{p+1}\phi^{p+1-t}||U_t-U_*||^2_F + \sum_{t=1}^{p+1}\phi^{p+1-t} ||V_t-V_*||^2_F.
\end{equation}
Once the algorithm converges, we obtain a new estimate of $V_*$ and this process continues.

This method captures the temporal patterns of the network in the low-dimensional representations $U_*$ and $V_*$. Additionally, this method can predict a link for any edge pair, even if there has yet to be one. Yet, there are several challenges to using this method in our context. First, it is well-known that NMF is a non-convex optimization problem so we may end up with a local optimum. Additionally, we must chose $q$, the latent dimension space of the non-negative matrices and $\phi$, the attenuation factor.

\paragraph{Graph neural network:} We also consider a state-of-the-art deep learning graph neural network method for link prediction called EvolveGCN \citep{pareja2020evolvegcn}. The basic idea of this method is that for a given time $t$, this method performs a graph convolution step on $G_t$ and then the corresponding weights are updated in the temporal direction using a recurrent neural network (RNN). Specifically, let $A_t, H_t^{(\ell)}$ and $W_t^{(\ell)}$ be the adjacency, node embedding and weight matrices, respectively, at time $t\in\{1,\dots,p\}$ for layer $\ell\in\{1,\dots,L\}$. Then the node embeddings for time $t$ are updated using a graph convolution step, i.e., 
\begin{equation}
    H_t^{(\ell+1)}
    =\text{GCONV}(A_t, H_t^{(\ell)}, W_t^{(\ell)}).
\end{equation}
The initial embedding matrix $H_t^{(0)}$ is the node features at time $t$ and the GCONV function simply normalizes the adjacency matrix before multiplying the normalized adjacency matrix with the other two inputs. Next, the authors propose two ways to temporally update the weight matrices for each graph convolution layer. The -H version treats $W_t^{(\ell)}$ as the hidden state of a dynamical system and updates the weights using the current node embeddings $H_t^{(\ell)}$ via a gated recurrent unit. The -O approach ignores the node embeddings and instead updates the weights using a long short term memory (LSTM) cell. These two steps together make up the EvolveGCN framework. 

This approach can easily be leveraged for link prediction. In particular, if $h_p^i$ and $h_p^j$ are, respectively, the $i$th and $j$th rows of the final embedding matrix $H_p$, then we can take their dot product to get a similarity score $S_{ij}$ for the likelihood of a link between the nodes $i$ and $j$, i.e., $S_{ij}=(h_p^i)^Th_p^j$. The larger $S_{ij}$, the more we expect an edge to exist between nodes $i$ and $j$ at time $p+1$, similar to the NMF approach. As such, this method can predict a link for any edge pair, regardless of whether there has been a historical link. Note that negative sampling and cross-entropy loss function are used to optimize the weights. In order to extend this method for predicting multiple time steps ahead, we first predict the state of the network one time step ahead, $\hat G_{p+1}$. Then this network can be fed into the fitted model to predict the ensuing time step $\hat G_{p+1}$ and so on.

There are several practical considerations for using this method for the IM problem.
First, this method requires node attributes. If these are unavailable, we compute the node2vec \citep{grover2016node2vec} embedding for each node at each snapshot and then use the output as the node feature. Since we believe that the graph structure plays a bigger role in the evolution of the network rather than these node features, we choose the -O version, per the authors' suggestion. This method scales well for large networks and can capture complicated dynamics driving link formation. This method, however, relies on negative sampling which is an important open problem in GNNs \citep[e.g.][]{robinson2020contrastive}. It also requires node features and it is not clear if node2vec is the optimal choice in the absence of meaningful domain features.

\paragraph{From output to predicted edges:}
Each link prediction method yields either a probability or similarity score for each edge pair. But this still leaves the important step of converting these continuous outcomes to a binary prediction of ``link'' or ``no link.'' To the knowledge of the authors, using link prediction methods for downstream tasks has received relatively little attention in the literature so this is a non-trivial step. The approach that we adopt preserves the ``average'' edge density in the network. Specifically, let $\hat \rho_t$ be the average probability of an edge for snapshot $t$ for $t=1,\dots,p$, i.e.,
\begin{equation}
\hat\rho_t=\sum_{i<j}\frac{A_{ij}^{(t)}}{{n \choose 2}}.
\end{equation}
Instead of taking the average of each, we give a larger weight to the more recent values. Then the weighted, average edge density is
\begin{equation}
    \hat\rho^*=\frac1{\sum_{t=1}^p\xi^{p-t}}\sum_{t=1}^p \xi^{p-t}\hat\rho_t
\end{equation}
where $0\leq\xi\leq1$. Thus, for each method, we select the top ${n\choose2}\hat\rho^*$ edge pairs and predict that they will have a link at the next time step. This also ensures that each method predicts the same number of edges for each future time step.

\subsection{Influence maximization:}
Once we have predicted the future evolution of the network, the second step of the proposed method is the IM algorithm. While there exists many temporal IM algorithms \citep[e.g.,][]{michalski2014seed, osawa2015selecting, erkol2020influence}, we consider a greedy algorithm and the dynamic degree discount algorithm \citep{murata2018extended}.

\paragraph{Greedy:}
The first algorithm that we consider is based on a greedy heuristic. At each step, a node is added to the seed set which results in the greatest expected marginal gain in influence spread (Algorithm \ref{alg:greedy} ). \cite{kempe2003maximizing} prove that for the IC and LT static propagation model, the greedy algorithm yields a solution within a factor of $(1-1/e)$ of the optimal spread so it is considered the ``gold standard'' in IM problems. This method also trivially accommodates any diffusion process. Since it requires computing the expected influence spread for $O(n)$ nodes at each of $k$ steps, however, the algorithm is computationally intensive and only feasible on small networks.

\begin{algorithm}[H]
\SetAlgoLined
\KwResult{Seed set $S$}
 {\bf Input: } Temporal network $\mathcal G$, seed size $k$\;
 
$S= \O$\;
 
 \For{$i \text{ in } 1:k$}{

 $v = \arg\max_{u\in V\setminus S}\{\sigma(S\cup\{u\}) - \sigma(S)\}$\;  
 
 $S = S\cup \{v\}$\;
 
  }
 \caption{Greedy influence maximization}
 \label{alg:greedy}
\end{algorithm}

\paragraph{Dynamic degree discount:}
In order to address the computational complexity of the greedy algorithm, \cite{murata2018extended} propose an IM heuristic algorithm based on dynamic degrees. First, the authors define the dynamic degree, $D_T(v)$, of node $v$ for a temporal network with $T$ snapshots as
\begin{equation}
    D_T(v)
    =\sum_{t=2}^T \frac{|N_{v,t-1}\setminus N_{v,t}|}{|N_{v,t-1}\cup N_{v,t}|}|N_{v,t}|
\end{equation}
where $N_{v,t}$ is the set of neighbors of node $v$ at time $t$, i.e., $N_{v,t}=\{u\in V:A_{uv}^{(t)}=1\}$. Then they extend the Degree Discount algorithm for static networks by \cite{chen2009efficient} to the temporal case. This method chooses the $k$ nodes with largest dynamic degree where the effect of selected nodes is removed or ``discounted'' from the remaining nodes. See Algorithm \ref{alg:deg} for details. \cite{murata2018extended} show that this method yields comparable results with the greedy algorithm but at a fraction of the run time.

\begin{algorithm}[H]
\SetAlgoLined
\KwResult{Seed set $S$}
 {\bf Input: } Temporal network $\mathcal G$, seed size $k$, infection parameter $\lambda$\;
 
$S= \O$ \;

$N_v = \bigcup_{t=1}^T N_{v,t}$\;

$dd_u=D_T(u)$ and $t_u=0$\;
 
 \For{$i \text{ in } 1:k$}{

 $v = \arg\max_{u\in V\setminus S}\{dd_u\}$\;  
 
 $S = S\cup \{v\}$\;

 \For{$u\in N_v$}{

    $t_u = t_u + 1$ \;

    $dd_u = D_T(u) - 2t_u - \{D_T(u)-t_u\}t_u\lambda$ \;
 
 }
 
  }
 \caption{Dynamic Degree Discount}
 \label{alg:deg}
\end{algorithm}

\subsection{Link prediction output heuristic}\label{sec:heur}

Thus far, we have laid out the path to combine link prediction algorithm with IM to find the seed nodes which maximize the influence on the unobserved future evolution of the network. Since we are primarily focused on IM, we are not strictly interested in the future evolution of the network, so much as in determining which nodes are the most ``important'' in the future. Thus, predicting the exact future evolution is not completely necessary. Additionally, we have seen how it is non-trivial to choose a cutoff to determine which edges to predict for the future and any sort of threshold that we use to predict edges will inherently lose information. Therefore, we propose the following IM heuristics based on the fitted link prediction model to determine the most likely influential nodes in the future networks. The idea is that if a node is likely to have edges with many other nodes, then it is also likely to be a good candidate for the IM seed set. Thus, if $\hat P$ is an $n\times n$ matrix returned by a link prediction algorithm where $\hat P_{ij}$ is the probability of an edge or similarity between nodes $i$ and $j$, the column sums of $P$ yield a useful measure of node importance for the IM task. Below, we describe the specific procedure for each method and include a general Algorithm in \ref{alg:lp}.

\begin{algorithm}[H]
\SetAlgoLined
\KwResult{Seed set $S$}
 {\bf Input: } Link prediction output $\hat P$, seed size $k$\;
 
Compute $\theta_i=\sum_{j=1}^n \hat P_{ij}$. \;

$S = \{\theta_{(i)}:i\leq k\}$ where $\theta_{(i)}$ is the $i$th order statistic. \;

 \caption{Link prediction IM}
 \label{alg:lp}
\end{algorithm}

\paragraph{LogRegSum:} The output of LogReg is the probability of a link for each edge pair that has previously been observed, i.e., $\hat p^{(p+1)}_{ij}:=\hat P_{ij}$ is the probability of an edge at time $t=p+1$ returned by the algorithm where $P_{ij}=0$ if $A_{ij}^{(t)}=0$ for $t\in\{1,\dots,p\}$. Thus, for each node $i$, we can sum the probabilities of node $i$ having an edge with every other node as a measure of importance, i.e., $\theta_i=\sum_j \hat P^{(p+1)}$. Then the $k$ nodes with largest value of $\theta_i$ are selected as the seed nodes.

\paragraph{NMFSum:} The output of NMF is the similarity score $\hat P_{ij}$ for each link $(i,j)$, computed as the cosine similarity between $(V_*)_i$ and $(V_*)_j$, i.e., $\hat P_{ij}=\cos((V_*)_i,(V_*)_j )$. The $k$ nodes with largest $\theta_i=\sum_j P_{ij}$ are again chosen as seed nodes.

\paragraph{GNNSum:} Once the GNN model is fit, the similarity score $\hat P_{ij}$ for each link $(i,j)$ is the dot product of $h_p^i$ and $h_p^j$ where $h_p^k$ is the $k$th row of the embedding matrix for time $t=p$, i.e., $\hat P_{ij}=(h_p^i)^T h_p^j$ and the seed nodes are chosen as in the other two cases.\\

There are several desirable features of these heuristics. First, they do not require a cutoff to predict future edges which is non-trivial. The lack of a threshold also means that we do not lose any information from the link prediction output; all of the information in the model is incorporated into the $\theta$ value. Similarly, we found in practice that often times the link prediction output will only have $n_p<n$ nodes which are active, i.e., have at least one predicted future edge. If $n_p<k$, then it is not clear how to chose the other $k-n_p$ nodes to include in the seed set. These heuristics, however, can always select $k$ seed nodes for any $k$. Another advantage is that we get the IM results for ``free'' after fitting the link prediction model, so it will be faster than any method that requires a second IM step, i.e., greedy or DynDeg. Lastly, the seed node selection does not depend on how far in the future we want to predict, nor the infection parameter $\lambda$, which is generally unknown in practice.

\section{Experiments}\label{sec:exp}
We perform experiments on one synthetic network and seven real-world networks from various domains in order to compare the different link prediction and IM algorithms. 

\subsection{Datasets}\label{sec:data}
We begin by briefly describing each dataset used in these experiments. See Table \ref{tab:data} for some relevant summary statistics. {\it Synthetic} is a synthetically generated network. For the first time step, an Erdos-Renyi \citep{erdos1959} graph was generated with $p=0.002$. Then for all subsequent time steps, 50\% of the previous edges were kept, and an equal amount were newly generated, thus preserving the total number of edges for each snapshot. All other networks are real-world networks. Several record an edge if two people are within close proximity: {\it Reality}, {\it High School 1}, {\it Hospital}, {\it Office} and {\it Copenhagen Bluetooth}. The remaining networks come from online interactions: emails in {\it Email4} and communication on a social media platform in {\it College}. These networks were selected because they have a wide range of nodes and edges, comprise different network-generating mechanisms, and come from a variety of domains. We stress that we intentionally chose different network sizes $(n,m)$, numbers of aggregation layers $(T,p)$, seed sizes $(k)$ and infection parameters $\lambda$ in order to compare the methods across a wide range of settings. Note that $\lambda$ varies between networks in order to ensure a reasonable amount of influence spread, and to be able to discriminate between the different methods.

\begin{table}[]
    \centering
    \begin{tabular}{l|ccccccccccr}
         Dataset & $n$ & $m$ & $\bar p$ & $T(p)$ & $fNT$ & $fLT$ & $FNT$ & $FLT$ & $DA$\\\hline
         
         Synthetic & 500 & 5174 & 0.002 & 20(10) & 1.00 & 0.53 & 0.75 & 0.00 & 0.00 \\
         Reality & 64 & 722 & 0.024 & 24(20) & 0.97 & 0.39 & 0.52 & 0.00 & $-0.17$\\
         Email4  & 168 & 3250 & 0.038 & 39(30) & 0.98 & 0.88 & 0.78 & 0.12 & $-0.30$\\ 
         HS 1  & 312 & 2242 & 0.005 & 20(16) & 0.98 & 0.74 & 0.39 & 0.02 & 0.09\\
         Hospital & 75 & 1139 & 0.052 & 16(12) & 0.83 & 0.64 & 0.29 & 0.02 & $-0.18$\\
         Office & 92 & 755 & 0.042 & 7(6) & 0.98 & 0.69 & 0.63 & 0.05 & $-0.06$\\
         CopenB & 703 & 21,318 & 0.002 & 100(90) & 0.94 & 0.64 & 0.27 & 0.00 & 0.08\\
         College & 1899 & 13,838 & 0.000 & 50(40) & 0.93 & 0.92 & 0.02 & 0.00  & $-0.19$
    \end{tabular}
    \caption{Summary statistics and temporal network measures for datasets under consideration. $n$: number of nodes; $m$: number of unique links; $\bar p:$ average edge density across time steps; $T(p)$: number of time steps (number used for training time); $fNT$: fraction of nodes present at half the sampling time; $fLT$: fraction of unique links present at half the sampling time; $FNT$: fraction of nodes present in the first and last 5\% of the sampling time; $FLT$: fraction of unique links present in the first and last 5\% of the sampling time; $DA$: degree assortativity. }
    \label{tab:data}
\end{table}

\subsection{Methods}
For each method, we use the given algorithm to find the optimal seed set on the {\it predicted} future graphs using different link prediction approaches. Then, using these seed sets, we compute the influence spread on the {\it actual} future evolution of the network, taking the average results over 1000 MC samples. We seek to find the methods with largest influence spread.

\begin{itemize}
    \item Oracle: assumes the future evolution of the network is known when finding the seeds nodes at the IM step, i.e., finds the nodes which maximize the spread on $G_{p+1},\dots,G_T$.
    \item Static (last): uses $G_p$ to find the optimal seed nodes and, therefore, implicitly assumes that the network does not continue to evolve
    \item Static (mem): constructs $G^{\rm mem}$ where there is a link between every edge pair with at least one link in $G_1,\dots,G_p$. Mathematically, if $A^{(t)}$ is the adjacency matrix associated with $G_t$ and $A^{\rm mem}$ is for $G^{\rm mem}$, then $A_{ij}^{\rm mem}=1$ if $A_{ij}^{(t)}=1$ for any $t\in\{1,\dots,p\}$ and 0 otherwise and IM algorithms are implemented on $G^{\rm mem}$. 
    \item JC: a simple link prediction heuristic based on Jaccard coefficients (JCs) \citep{liben2003link}. Consider $G_p$ and let nodes $u$ and $v$ be such that there is no edge between them, i.e., $A_{uv}=0$. Then the JC of nodes $u$ and $v$ is $|N_u\cap N_v|/|N_u\cup N_v|$ where $N_i$ is the neighbors of node $i$. We find the JC for all edge pairs without a link at time $p$ and predict an edge at time $p+1$ for the node pairs corresponding to the largest 5\% of JCs. In order to preserve the density of the network, we also randomly remove 5\% of edges. Once we have obtained $\hat G_{p+1}$, then this process can be repeated in order to predict the evolution of the network multiple steps in the future.
    \item LogReg: Uses the logistic regression method with LASSO penalty. The optimal penalty parameter is chosen from a grid search based on the best validation AUC using 75\% of the network time steps for training. The cutoff is chosen to preserve $\hat \rho^*$, the average network sparsity. 
    \item LogRegSum: Top $k$ nodes chosen as seed nodes based on LogReg heuristic described in Section \ref{sec:heur}
    \item NMF: Non-negative matrix factorization method. The dimension of $U,V$ is $0.05n$, i.e., five percent of the number of nodes. The algorithm is run 25 times with random initialization and the results with the lowest loss function are kept. The cutoff is chosen to preserve $\hat \rho^*$.
    \item NMFSum: Top $k$ nodes chosen as seed nodes based on NMF heuristic described in Section \ref{sec:heur} 
    \item GNN: EvolveGCN method. Node features are constructed using $d=16$ dimension node2vec embeddings. The model is trained over 200 epochs and the cutoff is chosen to preserve $\hat \rho^*$.
    \item GNNSum: Top $k$ nodes chosen as seed nodes based on GNN heuristic described in Section \ref{sec:heur}
\end{itemize}

Note that the dynamic degree algorithm does not apply to static networks---Static (last) and Static (mem)---so we apply a simple algorithm based on degrees. Please see the appendix for details.

\subsection{Results}

\subsubsection{Synthetic}
First, we generate a synthetic network using the process described in Section \ref{sec:data}. This network has $n=500$ nodes and $m=5,174$ unique edges. There are $T=20$ time steps with the first $p=10$ used for training and the last $T-p=10$ used for prediction. We fix $\lambda=0.05$ and vary the size of the seed set. The results are in Figure \ref{fig:synthetic_k} where we implemented the Dynamic Degree algorithm. LogReg yields the largest influence spread while NMFSum and GNNsum yield the smallest, but all methods are approximately the same. The total influence spread roughly increases linearly with $k$ for each method. When $k=50$, the influence spread from LogReg is about 87\% of that of Oracle. Note that the link prediction task for this network is extremely challenging as the process of adding and removing nodes is random and the time horizon for prediction is long compared with the number of training time steps.

\subsubsection{Reality}
\begin{figure}
    \centering
    \begin{subfigure}[b]{0.95\textwidth}
         \centering
         \includegraphics[width=\textwidth]{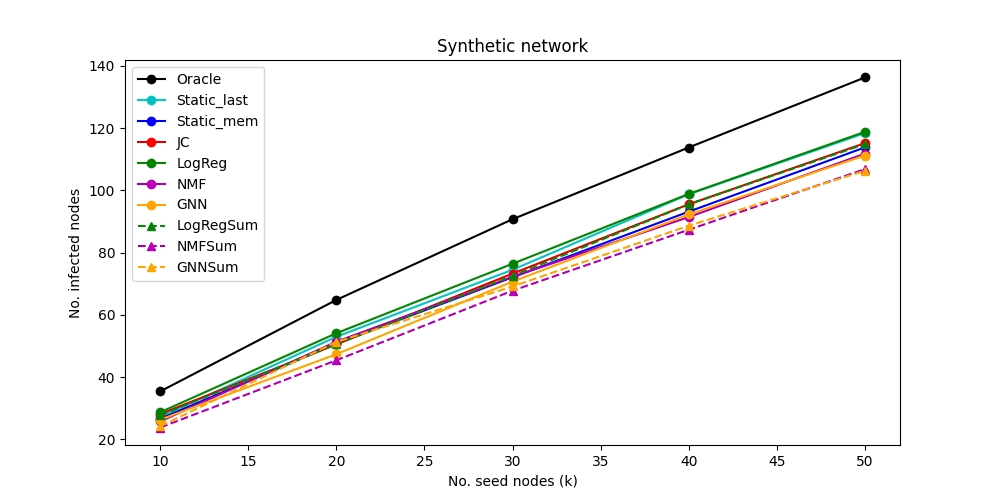}
         \caption{Synthetic (Dynamic Degree algorithm)}
         \label{fig:synthetic_k}
     \end{subfigure}
     \begin{subfigure}[b]{0.95\textwidth}
         \centering
         \includegraphics[width=\textwidth]{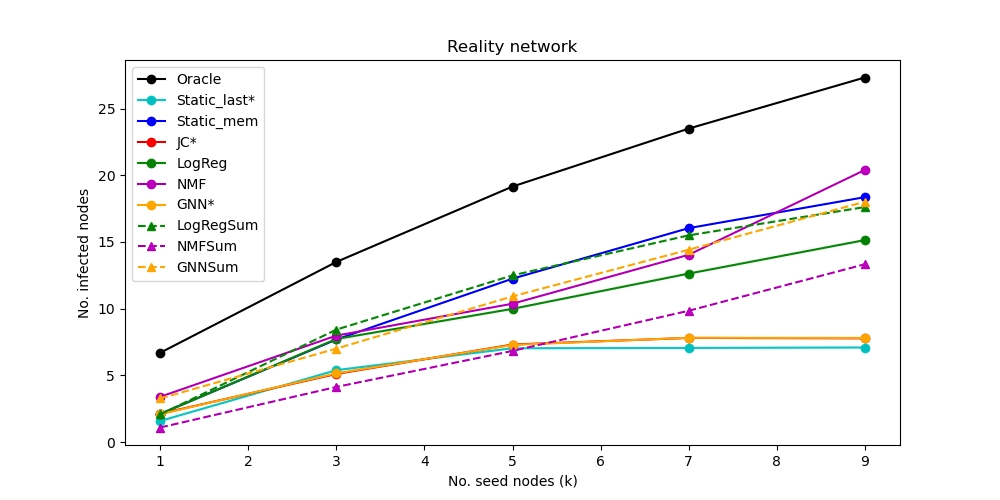}
         \caption{Reality (Greedy algorithm)}
         \label{fig:reality_k}
     \end{subfigure}
     \label{fig:email4}
     \caption{Influence maximization results on Synthetic and Reality networks.}
\end{figure}

The first real-world data set is the Reality network from \cite{eagle2006reality}. This network has $n=64$ nodes and $m=26,260$ edges where links were recorded every 5 seconds over an 8.63 hour period. We aggregate the network into evenly-spaced snapshots $G_1,\dots,G_T$ where $T=24$. We fix $\lambda=0.10, p=20, T-p=4$ and vary the size of the seed set. The results are in Figure \ref{fig:reality_k} using the Greedy algorithm. Note that Static (last), JC and GNN have a $^*$ because each method has less than $k$ active nodes for large $k$. Static (mem) and LogRegSum perform the best for almost all $k$ and come within about 67\% of the influence spread of Oracle. Both LogRegSum and GNNSum perform substantially better than LogReg and GNN, respectively, whereas NMFSum fares much worse than NMF. Static (last) and JC do not perform well in this scenario.

\subsubsection{Email}
Next, we consider an Email network (email4) from \cite{michalski2011matching}. This network has $n=167$ nodes and $m=82,927$ edges collected over 271 days with a granularity of 1 second. We aggregate the network into roughly one-week snapshots $G_1,\dots,G_T$ where $T=39$. Then we consider the first $p=30$ for training the link prediction algorithms and compare the results on the remaining $T-p=9$ graphs. For this network, we fix $\lambda=0.05$, vary the seed size $k$ and use the Dynamic Degree algorithm. The results are in Figure \ref{fig:email4_k}. Several methods come within 98\% of the performance of Oracle including LogReg, LogRegSum and GNNSum. For $k\geq 15$, Static (mem) yields the largest influence spread of the proposed methods while Static (last) and JC do well for small $k$. NMF and NMFSum perform the worst of all methods. LogRegSum performs about the same as LogReg and GNNSum also outperforms GNN. NMF and NMFsum are roughly the same.

\subsubsection{High school}
\begin{figure}
    \centering
     \begin{subfigure}[b]{0.95\textwidth}
         \centering
         \includegraphics[width=\textwidth]{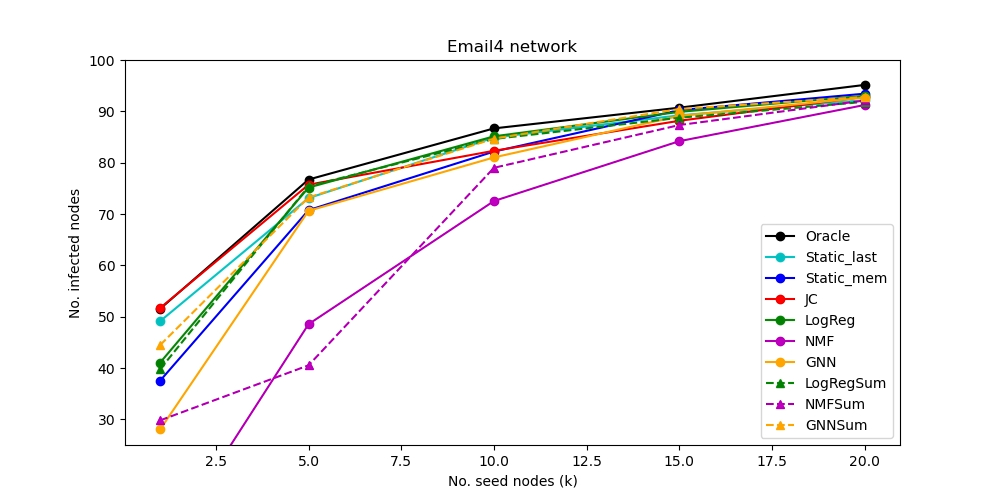}
         \caption{Email 4 (Dynamic Degree algorithm)}
         \label{fig:email4_k}
     \end{subfigure}
     \\
     \begin{subfigure}[b]{0.95\textwidth}
         \centering
         \includegraphics[width=\textwidth]{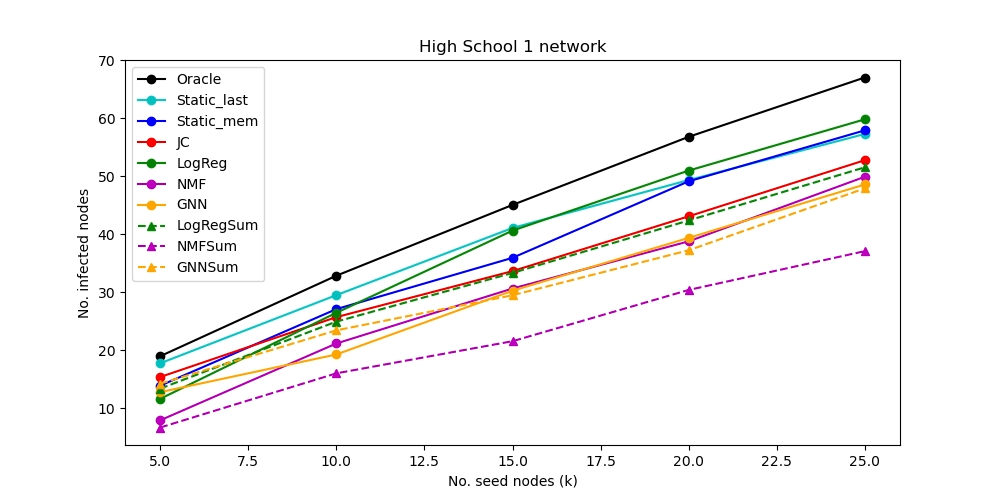}
         \caption{High school 1 (Dynamic Degree algorithm)}
         \label{fig:highschool1_k}
     \end{subfigure}
     \label{fig:email4}
     \caption{Influence maximization results on Email4 and High School 1 networks.}
\end{figure}

The High School 1 network comes from \cite{mastrandrea2015contact}. This network has $n=312$ nodes and $m=2242$ unique edges collected at 20 second intervals over 5 hours. We aggregate the network into roughly 15-minute snapshots $G_1,\dots,G_T$ where $T=20$. Then we consider the first $p=16$ for training the link prediction algorithms and compare the results on the remaining $T-p=4$ graphs. We vary the seed size $k$ and use the Dynamic Degree algorithm, fixing $\lambda=0.10$. The results are in Figure \ref{fig:highschool1_k}. Static (last) is the best method for $k\leq 15$ while LogReg yields the largest influence spread for large $k$. When $k=20$, LogReg achieves 90\% of the influence of Oracle. This is the only setting where LogReg outperforms LogRegSum. Static (mem) also performs well for large $k$ while GNN, NMF, GNNSum and NMFSum all struggle.

\subsubsection{Office}

The Office network comes from \cite{genois2015data} and consists of $n=92$ nodes and $m=755$ unique edges. There are no links in the temporal middle of this dataset, which is presumably the weekend when no workers interact at the office. Dropping these times, we aggregate the data into $T=7$ snapshots, representing one work day and predict the evolution of the network on the last day $(T-p=1$). We use the Greedy algorithm for IM and the results are in Figure \ref{fig:office_k} with $\lambda=0.10$. Since we are only predicting one snapshot into the future, Static (last) and JC yield the largest influence spread and come within 90\% of Oracle. Static (mem), GNNSum and LogRegSum also perform well here. GNNSum outperforms GNN for all $k$ and LogRegSum fares better than LogReg for most $k$. Again, NMF and NMFSum yield the lowest influence spread.

\subsubsection{Hospital}
\begin{figure}
    \centering
    \begin{subfigure}[b]{0.95\textwidth}
         \centering
         \includegraphics[width=\textwidth]{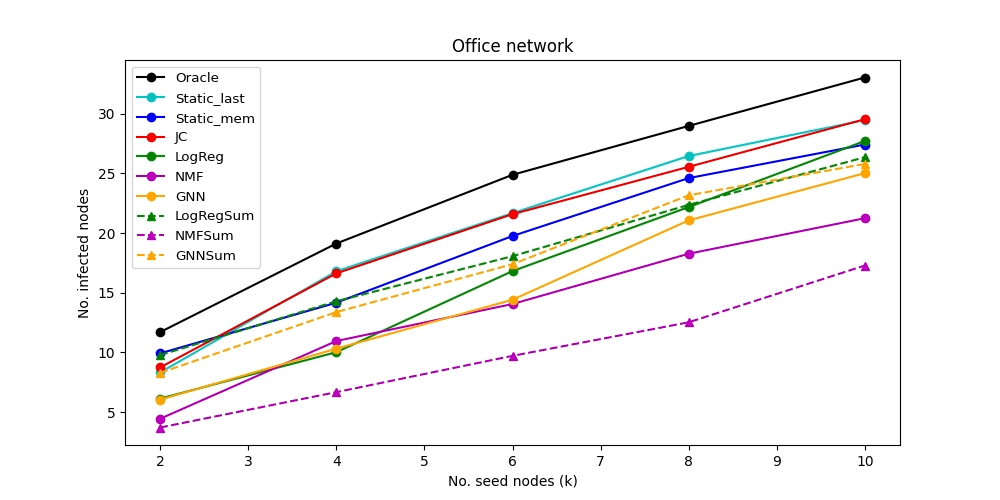}
         \caption{Office (Greedy algorithm)}
         \label{fig:office_k}
     \end{subfigure}
     \begin{subfigure}[b]{0.95\textwidth}
         \centering
         \includegraphics[width=\textwidth]{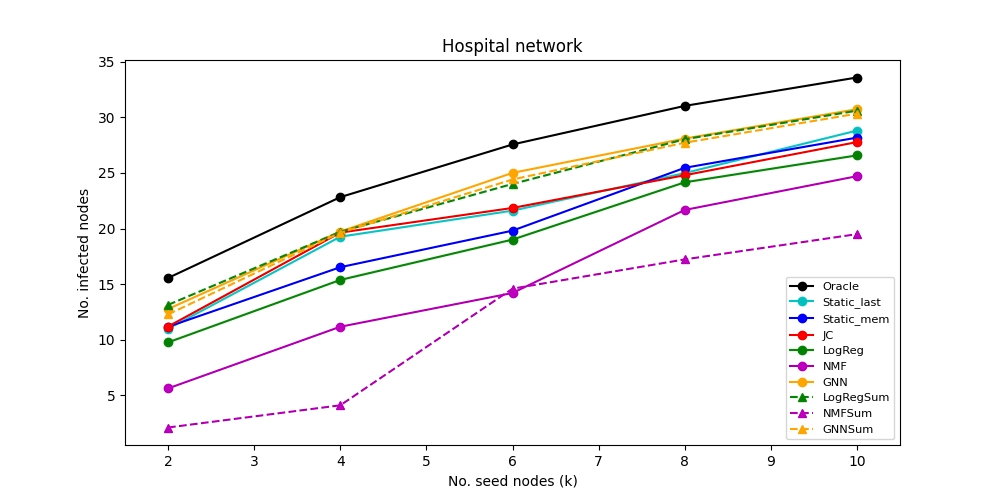}
         \caption{Hospital (Greedy algorithm)}
         \label{fig:hosp_k}
     \end{subfigure}
     \label{fig:email4}
     \caption{Influence maximization results on Office and Hospital networks.}
\end{figure}

With $n=75$ nodes and $m=1139$ unique links, the hospital network \citep{vanhems2013estimating} is the next that we consider. We aggregate the network at six hour intervals, yielding $T=16$ snapshots and we use the first $p=12$ for training the models and the final day for prediction. The results are in Figure \ref{fig:hosp_k} using the Greedy algorithm and $\lambda=0.10$. GNN, GNNSum and LogRegSum yield the largest influence spread which is about 91\% of that of Oracle. LogRegSum also does much better than LogReg. Static (last) and JC perform better than Static (mem) for small $k$ and about the same for $k\geq 8$. Both NMF methods have the lowest spread of influence.

\subsubsection{Copenhagen Bluetooth}
\cite{sapiezynski2019interaction} collected the Copenhagen Bluetooth network with $n=703$ nodes and $m=21,318$ unique edges. Data was collected over several weeks at 5 minute intervals. We aggregate the network at $T=100$ evenly spaced intervals and use the first $p=90$ for training the model. We set $\lambda=0.05$, vary the seed size $k$ and compute the optimal seed nodes using the Dynamic Degree algorithm.
The results are in Figure \ref{fig:cpbt_k}. Static (mem) yields the largest influence spread and comes within 99\% of the spread of Oracle. LogRegSum has almost the same influence as Static (mem), especially for large $k$. Part of the reason that Static (last) does not perform as well for large $k$ is that it has less than 150 active nodes, similar to GNN and JC. NMF and NMFSum perform similarly and are comparable to GNNSum when $k\geq 100$.

\subsubsection{College} 
\begin{figure}
    \centering
    \begin{subfigure}[b]{0.95\textwidth}
         \centering
         \includegraphics[width=\textwidth]{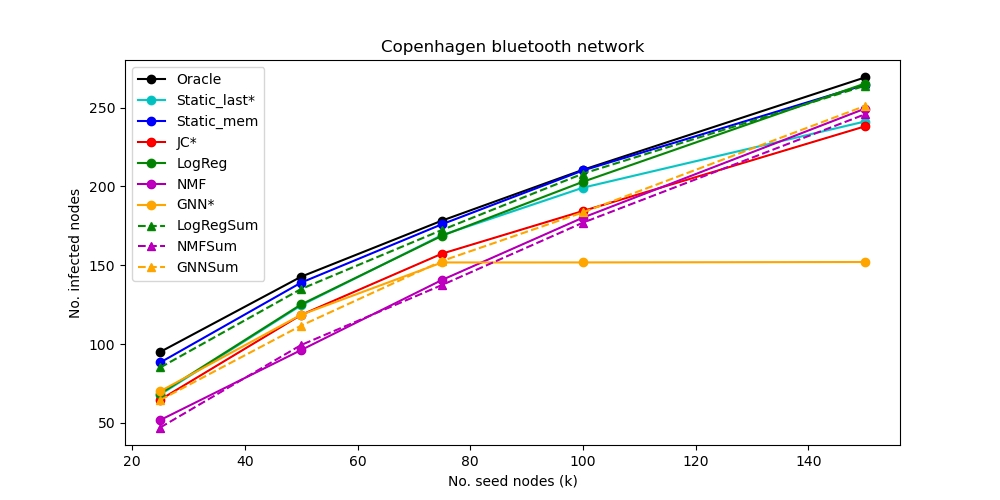}
         \caption{Copenhagen Bluetooth (Dynamic Degree algorithm)}
         \label{fig:cpbt_k}
     \end{subfigure}
     \begin{subfigure}[b]{0.95\textwidth}
         \centering
         \includegraphics[width=\textwidth]{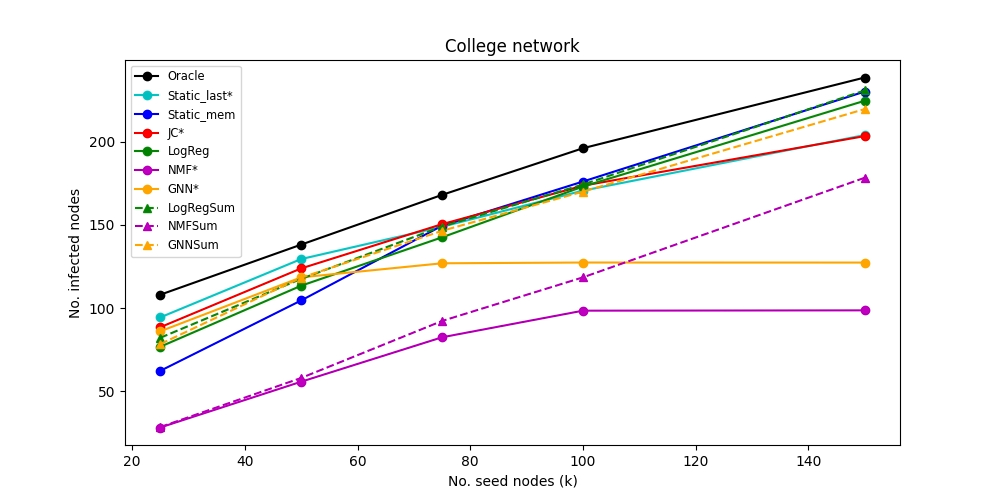}
         \caption{College (Dynamic Degree algorithm)}
         \label{fig:college_k}
     \end{subfigure}
     \label{fig:email4}
     \caption{Influence maximization results on Copenhagen Bluetooth and College networks.}
\end{figure}

The college network \cite{panzarasa2009patterns} has $m=13,838$ unique edges among $n=1899$ nodes. Links are recorded at 1 second intervals over the 193 days. We aggregate the network into $T=50$ snapshots over equally spaced intervals and hold out the last $T-p=10$ snapshots for prediction. We fix $\lambda=0.25$ and vary $k$, using the Dynamic Degree algorithm to find the optimal seed nodes. The results are in Figure \ref{fig:college_k}. For smaller $k$, Static (last) and JC yield the greatest influence. For larger $k$, however, Static (mem), LogReg and LogRegSum do the best and come within 96\% of Oracle. Static (last) and JC are again hampered by having less than $k$ active nodes for large $k$. NMF and NFMSum continue to yield the smallest influence spread.

\subsection{Discussion}

There are several interests trends that emerge from these experiments. First, we observed that for every data set, save Reality, the best performing method came within 87\% of the influence spread of the Oracle method, and in one case (Copenhagen Bluetooth), was as high as 99\%. This is a promising finding as it shows that meaningful seed nodes can be found for IM, even when the future evolution of the network is unobserved. Of the proposed methods, LogRegSum proved itself to be the best method as it yielded the largest influence spread for many settings, followed by GNN/GNNSum, while NMF/NMFSum regularly yielded the smallest influence. Perhaps the most interesting finding, however, is that the simple heuristic based solely on historical edges, Static (mem), consistently yielded one of the best performances. These results held across networks with different sizes, temporal characteristics, prediction duration, and IM algorithms. Because Static (mem) does not have a link prediction step, it is substantially faster than all of the proposed methods (including LogRegSum), so we suggest this method for use in practice.

There are also some noteworthy connections between the temporal network statistics from Table \ref{tab:data} and the IM results. For example, in the Email 4 network, all methods may have performed well because this network has less temporal variation. $fNT$, $fLT$, $FNT$ and $FLT$ are all higher in this data set than for other networks, indicating there might be less temporal variation. For instance, $FNT$ being large means that the majority of nodes are present at the beginning and end of the network's life cycle (and presumably in-between as well). Another observation is that the GNN methods do comparably worse in networks with a strong community structure (High School 1 and Copenhagen Bluetooth), as measured by degree associativity. Interestingly, even though the College network has very few nodes and edges in the first and last 5\% of the sampling time which indicates significant temporal variation, several methods still perform well. Static (mem) also performs well on sparser networks (High School 1, Copenhagen and College). Having fewer edges in the network likely makes the link prediction task more difficult, so aggregating across time turns out to be the best way to illuminate a node's influence. In the Appendix, we also report the true positive rate (TPR) and relative degree mean square error (MSE) for JC, LogReg, NMF and GNN on each network which quantifies the quality of the link predictions. These results highlight the difficulty of link prediction as a TPR greater than 0.40 is never achieved, nor is an MSE below 0.60.

The strong performance of LogRegSum and Static (mem) yields some key insights into the {\it ex ante} IM task on temporal networks. It first demonstrates that predicting the future evolution of the network is not strictly necessary to determine the most influential seed nodes. Instead, the majority of the information needed to select seed nodes can be extracted from the previous edge history while the actual evolution of the network does little to change the importance of nodes. This unexpected result---somewhat in disagreement with results from the related, yet distinct, vaccination problem  \cite{holme_vacc}---indicates that while a network is evolving over time, the underlying importance of nodes, from an IM-sense, may not change. Thus, elucidating the influential nodes from the observed network is more important than predicting the future evolution. Indeed, link predictions may yield noisy results and inherently lose information due to cutoffs and thresholds, so methods that only consider historical data, such as LogRegSum and Static (mem), may perform better by ``averaging out'' some of this noise. Additionally, in light of Static (mem) performing so well, it should be no surprise that LogRegSum also does well since this method only predicts links for edge pairs with at least one historical edge. Thus, even though GNN and NMF can predict links for any edge pairs, this may not be an advantageous feature. We stress that these findings only apply for the SI model, and may not be applicable under other diffusion mechanisms. Lastly, we saw that if a link prediction method or heuristic yielded predictions with $n_p<k$ active nodes, then choosing the remaining $k-n_p$ seed nodes becomes a non-trivial task. This problem is avoided with methods such as LogRegSum and Static (mem), which always allow for all $n$ nodes to be considered for the seed set. We stress that these conclusions only apply to these datasets and the SI diffusion model.

\section{Conclusions}\label{sec:conc}

In this work, we addressed the important problem of {\it ex ante} influential maximization on temporally-varying networks. We first predict the future evolution of the network and then use a standard temporal IM algorithm on the predicted network to find the optimal seed nodes. We also proposed IM heuristics using the model fit of the link predictions to find seed nodes, omitting the actual link prediction step. Across many settings, we demonstrated influence spread using the proposed methods on par with the gold standard method of comparison, with LogRegSum performing the best. These results show that it is possible to construct satisfactory seed sets for the IM task, even when the future topology of the network is unknown. Additionally, we found that in many cases, a simple heuristic based on historical edges yielded the best results and in practice we suggest this method due to its performance, simplicity and computational advantage. Our surprising results indicate that the most influential nodes may not vary with time, even as the network topology does.

We emphasize that these results were shown under the SI diffusion model. It is possible that under a different model, e.g., IC or LT, Static (mem) may not perform as well. Indeed, since the SI model allows nodes to activate their neighbors at any time following infection (nodes never become inactive), it is possible that the timing of when links occur is less important than the absence or presence of a link at any time. For example, \cite{gayraud2015diffusion} and \cite{michalski2014seed} both highlight the importance of the timing of links and that the performance of their methods suffers when all snapshots are simply aggregated, but this was for the IC and LT models. Even when using the SIR model, a close cousin of the SI model, \cite{erkol2020influence} show that an approach based on aggregated temporal snapshots performed poorly. That being said, Gayraud and Erkol's results were shown under the {\it ex post} assumption. Regardless, in practice it is paramount to understand the diffusion mechanism for the particular problem at hand as the performance of these algorithms may vary.

There are several interesting avenues of future work. Using link prediction methods for down-stream tasks has received relatively little attention and we have highlighted several challenges. We proposed a heuristic to side-step this challenge and it would be interesting to study other {\it ex ante} tasks and see if the same results hold. Additionally, any link prediction method inherently has uncertainty in its outputted edges. Another direction of future work is determining how to incorporate this uncertainty into the IM task. LogReg would be a sensible method to work with since it yields probabilities of edges, as compared to NMF or GNN which only yield similarity scores.

\section*{Acknowledgements} The authors would like to thank the editors and anonymous reviewers whose comments greatly improved the quality of the manuscript.

\section*{Data availability} All data used in this work are freely available online.
 
\section*{Funding}
This work was conducted while EY was on a fellowship funded by the Japan Society for the Promotion of Science (JSPS). PH was supported by JSPS KAKENHI Grant Number JP 21H04595.

\section*{Authors contributions}
EY was responsible for the conception, design, analysis, interpretation and writing of this manuscript. TM verified the results, assisted in interpretation of results and provided mentorship and editing. PH provided the data, assisted in interpretation of results and provided mentorship and editing.

\bibliographystyle{apalike}
\bibliography{refs}

\clearpage

\section*{Appendix}
\subsection*{Degree algorithm for static networks}

IM algorithm used for Static (last) and Static (mem) for all simulation settings where Dynamic Degree Algorithm was used. Note that we could have directly used the algorithm from \cite{chen2009efficient}, but we found that the algorithm below yielded slightly better performance in practice. 

\begin{algorithm}[H]
\SetAlgoLined
\KwResult{Seed set $S$}
 {\bf Input: } Adjacency matrix $A$, seed size $k$\;
 
$S= \O$ \;

Compute degrees: $d_u = \sum_{v=1}^n A_{uv}$ \;

Find neighbors: $N_u = \{v:A_{uv}=1\}$

 \For{$i \text{ in } 1:k$}{

 $v = \arg\max_{u\in V\setminus S}\{d_u\}$\;  
 
 $S = S\cup \{v\}$\;

 \For{$u\in N_v$}{

    $d_u = d_u - 1$ \;
 
 }
 
  }
 \caption{Dynamic Degree Discount}
 \label{alg:deg}
\end{algorithm}

\subsection*{Link prediction metrics}

In Table \ref{tab:linkpred}, we report metrics which quantify the quality of the link prediction algorithms. We report the true positive rate (TPR) and relative degree root mean squared error (MSE) for each method and each network. TPR is the proportion of future edges that the method correctly predicted averaged over all time steps where larger is better. MSE is the root mean squared error between the aggregated degrees of the future and predicted networks, divided by the future average degree. Mathematically, if ${\bf d}_0$ and ${\bf d}_1$ are the true and estimated aggregated degree sequence of $G_{p+1},\dots,G_T$, respectively, then
$$
    MSE
    =\sqrt{\frac{1}{n}\sum_{i=1}^n \left(\frac{d_{0i}-d_{1i}}{d_{0i}}\right)^2}.
$$
Here, smaller is better. The TPR is particularly low for each method for Synthetic, Reality and High School 1, which also corresponds to some of the weaker performances of the respective methods. Some of the best results occur for Email 4, which also corresponds to some of the largest influence spread results compared with Oracle. These results show how difficult link prediction is as a TPR greater than 0.40 is never achieved, nor is an MSE below 0.60.

\begin{table}[]
    \centering
    \begin{tabular}{ll|cc}
        Dataset & Method & TPR & MSE  \\\hline
        Synthetic 
        & JC & 0.08 & 0.65  \\
        & LogReg & 0.04 & 0.58  \\
        & NMF & 0.00 & 1.22  \\
        & GNN & 0.01 & 2.24   \\\hline
        Reality 
        & JC & 0.02 & 1.26 \\
        & LogReg & 0.04 & 1.19 \\
        & NMF & 0.04 & 0.99 \\
        & GNN & 0.04 & 1.28\\\hline
        Email 4 
        & JC & 0.39 & 0.65  \\
        & LogReg & 0.22 & 0.82  \\
        & NMF & 0.06 & 1.49  \\
        & GNN & 0.22 & 1.38 \\\hline   
        High School 1 
        & JC & 0.00 & 1.05  \\
        & LogReg & 0.00 & 1.02 \\
        & NMF & 0.08 & 1.64 \\
        & GNN & 0.05 & 1.67 \\\hline   
        Hospital 
        & JC & 0.36 & 2.38 \\
        & LogReg & 0.08 & 1.26  \\
        & NMF & 0.11 & 1.42  \\
        & GNN & 0.27 & 1.53 \\\hline   
        Office 
        & JC & 0.34 & 0.86  \\
        & LogReg & 0.20 & 0.79  \\
        & NMF & 0.16 & 1.19  \\
        & GNN & 0.15 & 1.26 \\\hline  
        CopenB 
        & JC & 0.05 & 2.16  \\
        & LogReg & 0.02 & 1.69  \\
        & NMF & 0.02 & 2.13  \\
        & GNN & 0.02 & 4.56 \\\hline   
        College 
        & JC & 0.14 & 7.25  \\
        & LogReg & 0.02 & 6.09  \\
        & NMF & 0.00 &  4.98 \\
        & GNN & 0.03 & 19.14 
    \end{tabular}
    \caption{True positive rate (TPR) and relative degree root mean squared error (MSE) for each method and each network. TPR is the proportion of future edges that the method correctly predicted averaged over all time steps (larger is better). MSE is the root mean squared error between the aggregated degrees of the future and predicted networks, divided by the future average degree (smaller is better).}
    \label{tab:linkpred}
\end{table}

\end{document}